%%%%%%%%%%%%%%%%%%%%%%%%%%%%%%%%%%%%%%%%%%%%%%%%%%%%%%%%%%%%%%
% Dynamical correlation functions in the Calogero-Sutherland %
% Model, F. Lesage, V. Pasquier, D. Serban.                  %
%                                                            %
%%%%%%%%%%%%%%%%%%%%%%%%%%%%%%%%%%%%%%%%%%%%%%%%%%%%%%%%%%%%%

\input harvmac.tex
\input epsf.tex
\overfullrule=0mm
\newcount\figno
\figno=0
\def\fig#1#2#3{
\par\begingroup\parindent=0pt\leftskip=1cm\rightskip=1cm\parindent=0pt
\baselineskip=11pt
\global\advance\figno by 1
\midinsert
\epsfxsize=#3
\centerline{\epsfbox{#2}}
{\bf Fig. \the\figno:} #1\par
\endinsert\endgroup\par
}
\def\figlabel#1{\xdef#1{\the\figno}}
\def\encadremath#1{\vbox{\hrule\hbox{\vrule\kern8pt\vbox{\kern8pt
\hbox{$\displaystyle #1$}\kern8pt}
\kern8pt\vrule}\hrule}}
%%%%%%%%%%%%%%%%%%%%%%%%%%%%%%%%%%%%%%%%%%%%%%%%%%%%%%%%%%%%%%%
%		%DEFINITIONS FOR TEX
%
\def\bar{\overline}
\def\({\left(}			
\def\){\right)}		
\def\[{\left[}		
\def\]{\right]}		
%
%%%%%%%%%%%%%%%%%%%%%%%%%%%%%%%%%%%%%%%%%%%%%%%%%%%%%%%%%%%%%%%
%
\def\frac#1#2{{#1 \over #2}}

\def\2pi{\hbox{$2\pi i$}}
%
%%%%%%%%%%%%%%%%%%%%GREEK LETTERS%%%%%%%%%%%%%%%%%%%%%%%%%%%%%%
%

\def\be{\beta}

\def\la{\lambda}

%
%%%%%%%%%%%%%%%%%%%%%%%%%%%%%%%%%%%%%%%%%%%%%%%%%%%%%%%%%%%%%%%%%%%%%%

%%%%%%%%%%%%%%%%%%%%%%%%%%%%%%%%%%%%%%%%%%%%%%%%%%%%%%%%%%%%%%%%%%%%%%%%%%%%
% References 
\lref\Macdo{I.G. Macdonald, S\'eminaire Lotharingien, Publ. I.R.M.A.
Strasbourg, 1988 .}
\lref\Haldcorr{F.D.M. Haldane, M.R. Zirnbauer, Phys. Rev. Lett. Vol. 71,
No. 24, 4055.}
\lref\Haldrev{F.D.M. Haldane, To appear in "Proceedings of the 16th
Taniguchi Symposium", 1993.}
\lref\alti{B.D. Simons, P.A. Lee and B.L. Altshuler, Nucl. Phys. B409,
(1993) 487-508.}
\lref\altii{B.D. Simons, P.A. Lee, and B.L. Altshuler, 
Phys. Rev. Lett. V.70, No26, (1993) 4122.}
\lref\calo{
F. Calogero, J. Math. Phys. {\bf 10}, 2191, (1969).}
\lref\suth{B. Sutherland, J. Math. Phys. {\bf 12} , 246 (1971);
{\bf 12} , 251 (1971).}
\lref\Haldconj{F.D.M. Haldane, To appear in proceedings of
the International Colloquium in Modern Field Theory, Tata 
institute, 1994.}
\lref\Forres{P. J. Forrester, Nucl. Phys. B388 (1992), 671-699.}
\lref\tables{E.R. Hanson, {\it A Table of Series and Products},
Prentice Hall, N.J., 1975.}
\lref\stanley{R. P. Stanley, Adv. in Math., 77, (1989) 76-115.}
\lref\Haldspin{F.D.M. Haldane, Phys.Rev.Lett. 66 (1991) 1529}
\lref\Haldstat{F.D.M. Haldane, Phys.Rev.Lett. 67 (1991) 937}
\lref\faddeev{L.D.Faddeev, L.A.Takhtajan, Phys.Lett. 85A (1981) 375 }
\lref\macdolivre{I.G. Macdonald, {\it Symmetric functions and Hall
polynomials}, Clarendon Press, (1979).}
\lref\Ha{Z.N.C.Ha, F.D.M.Haldane, Phys.Rev.D 47 (1993) 12459}
\lref\YB{D.Bernard, M.Gaudin, F.D.M.Haldane and V.Pasquier, J.Phys.A 26
(1993) 5219}
\lref\Gaudin{M.Gaudin, Saclay preprint SPhT/92-158.}
\lref\pm{J. A. Minahan, A. P. Polychronakos, hep-th/9404192 .}
\lref\ha{Z.N.C. Ha, Cond-mat 9405063.}
%%%%%%%%%%%%%%%%%%%%%%%%%%%%%%%%%%%%%%%%%%%%%%%%%%%%%%%%%%%%%%%%%%%%%%%%%%%%

\Title{SPhT/94-051}
{{\vbox {
%\centerline{}
\bigskip
\centerline{Dynamical correlation functions in the}
\centerline{Calogero-Sutherland model.} }}}
\bigskip
\centerline{F. Lesage, V. Pasquier and D. Serban}

\bigskip

\centerline{ \it Service de Physique Th\'eorique de Saclay
\footnote*{Laboratoire de la Direction des Sciences 
de la Mati\`ere du Commissariat \`a l'Energie Atomique.},}
\centerline{ \it F-91191 Gif sur Yvette Cedex, France}

\vskip .5in

We compute the dynamical Green function
and density-density correlation in the Calogero-Sutherland
model for all integer values of the coupling constant.  
An interpretation of the intermediate states in terms of
quasi-particles is found.

\noindent
\vskip 2.0in
{\noindent \it Submitted to Nuclear Physics B.}
\Date{04/94}

\newsec{Introduction}

The Calogero-Sutherland Hamiltonian was defined in \suth\
\calo\ as a model of particles on a circle 
interacting with a long range potential.  The positions
of the particles are denoted by $x_i$, $1\leq i \leq N$,
$0\leq x_i \leq L$.  The total momentum and
the hamiltonian which give their dynamics are respectively
given by:
\eqn\imp{\hat{P}=\sum_{j=1}^N {1\over i} {d\over dx_i}}
\eqn\halm{\hat{H}=
-\sum_{j=1}^N {1\over 2}{d^2 \over dx_i^2} + \beta (\beta-1)
{\pi^2\over L^2} \sum_{i<j} {1\over \sin^2(\pi (x_i-x_j)/L)}
}
$\beta$ is a positive constant which we shall take to be an 
integer.  The wave functions solutions of the equation $\hat{H} \Psi
=E\Psi$ have the following structure
\eqn\wavfct{
\Psi(x)=\Delta^\beta(x) \Phi(x)
}
where
\eqn\waveftci{
\Delta(x)=\prod_{i<j} \sin\left( {\pi (x_i-x_j) \over L}\right)
}
and $\Phi(x)$ denotes a polynomial in the variables $z_j=e^{i 2 \pi x_j/L}$
and $z_j^{-1}$ symmetric under the permutations of the indices.  
An eigenstate of $\hat{H}$ is
characterized by the highest weight monomial of $\Psi(x)$,
\eqn\hw{
\exp ( {2 \pi i\over L} \sum_{j=1}^N k_j x_j )
}
where the $k_j$'s define an increasing sequence of (not necessarily positive)
integers \foot{For simplicity we take $\beta (N-1)$ to be an even integer
so that the ground state is non-degenerate}
subject to the constraint
\eqn\constra{
k_{j+1}-k_j \geq \beta
}
The eigenvalues of $\hat{P}$ and $\hat{H}$ on this eigenstate
are given by:
\eqn\estate{\eqalign{
\hat{P}\vert k_1 ... k_N>&=\sum_{j=1}^N {2 \pi k_j\over L} \vert k_1 ... k_N>
\cr \hat{H} \vert k_1 ... k_N>&=\sum_{j=1}^N {1\over 2} \left(
{2 \pi k_j\over L} \right)^2 \vert k_1 ...  k_N >
}}
The wave function for these states will be described in detail in
the next section.

The results reported here are the expression of the retarded Green
function and the density-density dynamical correlation function
at arbitrary integer value of the coupling constant $\beta$.
Here, we give the result in the thermodynamic limit, $N\rightarrow
\infty$, $L\rightarrow \infty$, keeping the density
$\rho=N/L$ fixed.  The exact expressions for a finite number
of particles are given in section 3 and 4.

The retarded Green function is defined by:
\eqn\retGreen{\eqalign{
\eta& (x,t;x',t')=\ <0\vert \Psi^\dagger(x,t) \Psi(x',t')\vert 0> \cr
\noalign{\vskip3pt}
&={\cal N}^{-1} \rho \left( \prod_{k=1}^N\int_0^L dy_k\right)
 \prod_i z_i^{\beta
(N-1)/2} \Delta^\beta(x',y_k) 
e^{-i(\hat{H}-E_0)(t'-t)} \Delta^\beta (x,y_k) \prod_l z_l^{-\beta(N-1)/2}
}}
with 
\eqn\nr{
{\cal N}=\int_0^L dx \prod_{k=1}^N dy_k \Delta^{2\beta}(x,y_k)
}
and it is equal to:
\eqn\introfinal{
\eta(x,t,x',t')=  
{\rho \over 2} \left(
\prod_{i=1}^\beta  {\Gamma(1+1/\beta)\over
\Gamma(i/\beta)^2}\int_{-1}^1 dv_i \right) f_\beta(v_i)
 e^{i(Q(x-x')-E(t-t'))} 
}
\eqn\ffact{f_\beta(v_i)=
\prod_{i=1}^\beta \ (1-v_i^2)^{-1+1/\beta} \times 
 \prod_{i<j} \vert v_i-v_j\vert^{2\over \beta} 
}
The energy and momentum are given by
\eqn\EAii{
E={-\rho^2 \pi^2 \beta\over 2}\sum_{i=1}^\be v_i^2, \qquad 
Q=\pi \rho \sum_{i=1}^\be v_i }
The case of physical interest is for $\beta=2$, $\rho=1/2$, for which
$\eta$ gives the spin-spin correlation function of the 
Haldane-Shastry chain.  In this case, the spin operator
creates 2 holes which propagate freely with velocities $v_i$.

The density-density correlation function is given by~:
\eqn\dende{
g_\beta(x,t)=\ <0|\rho(x,t) \rho(0,0) |0>
}
where $\rho(x)=\sum_{i=1}^N \delta(x-x_i)$.
We have obtained the following expression in the thermodynamic limit~:
\eqn\introtherlim{\eqalign{
A(\beta) \int_1^\infty & dw\int_{-1}^1 dv_1 ... \int_{-1}^1 dv_\beta \
(w^2-1)^{\beta-1} \prod_{i=1}^\beta (1-v_i^2)^{-1+1/\beta} \times \cr
&\times \prod_{i<j} \vert v_i-v_j \vert^{2/\beta}  \times
{(\Sigma_i v_i-\beta w)^2 \over \prod_{i=1}^\beta (v_i-w)^2 }
 e^{-iEt} \cos (Q x)
 }
}
with the constant~:
\eqn\introconst{
A(\beta)={\rho^2\over 2 \beta} \prod_i
{\Gamma(1+1/\beta)\over \Gamma(i/\beta)^2 }
}
and the energy and momentum~:
\eqn\EAm{
E={\pi^2 \rho^2 \beta \over 2}
\left( \be w^2-\sum_{i=1}^\be v_i^2 \right) , \quad Q=\pi \rho
\left( \sum_{i=1}^\be v_i-\be w \right) .}
The physical interpretation of this result is that the density 
operator $\rho(x)$ creates excitations composed of $\beta$ holes
with velocity $v_i$ and a particle with velocity $w$ and mass $\be$ upon
acting on the vacuum.

The method followed here to compute a correlation function
consists in inserting a complete set of eigenstates in 
between the two operators evaluated at time $t$ and at 
time $0$.  
Before going to the precise calculation, let us determine the 
states which propagate and give their interpretation as quasiparticles.
In the absence of second quantization our argument is heuristic
but in complete agreement with the exact results.

Let us describe the ground state as a filled Fermi sea occupied with
particles of momenta $k_i$ subject to the constraint \constra .
Then, in the ground state we have~:
\eqn\fs{
k_i=\beta \(i-{N+1\over 2}\), \ \ \ 1\leq i \leq N
}
and the Fermi momentum is given by $k_F=\beta {N-1\over 2}$.
This is described in figure 1 for $\be =2$, $N=5$.
The ones stand for the occupied
momenta and the zeroes for the empty momenta.
Let us now consider the case of the Green function.
The action of the operator $\Psi(x)$ is to remove
a particle from the Fermi sea {\bf without changing the Fermi
level}. Therefore, the intermediate
states which can propagate are given by all the possible
arrangements of $N-1$ momenta $k_i$ inside the 
Fermi sea $(-k_F\le k_i \le k_F )$ and subject to the constraint \constra .
\fig{a) The ground state, b) an excitation created by $\Psi(x)$, c) an excitation created by $\rho(x)$.}{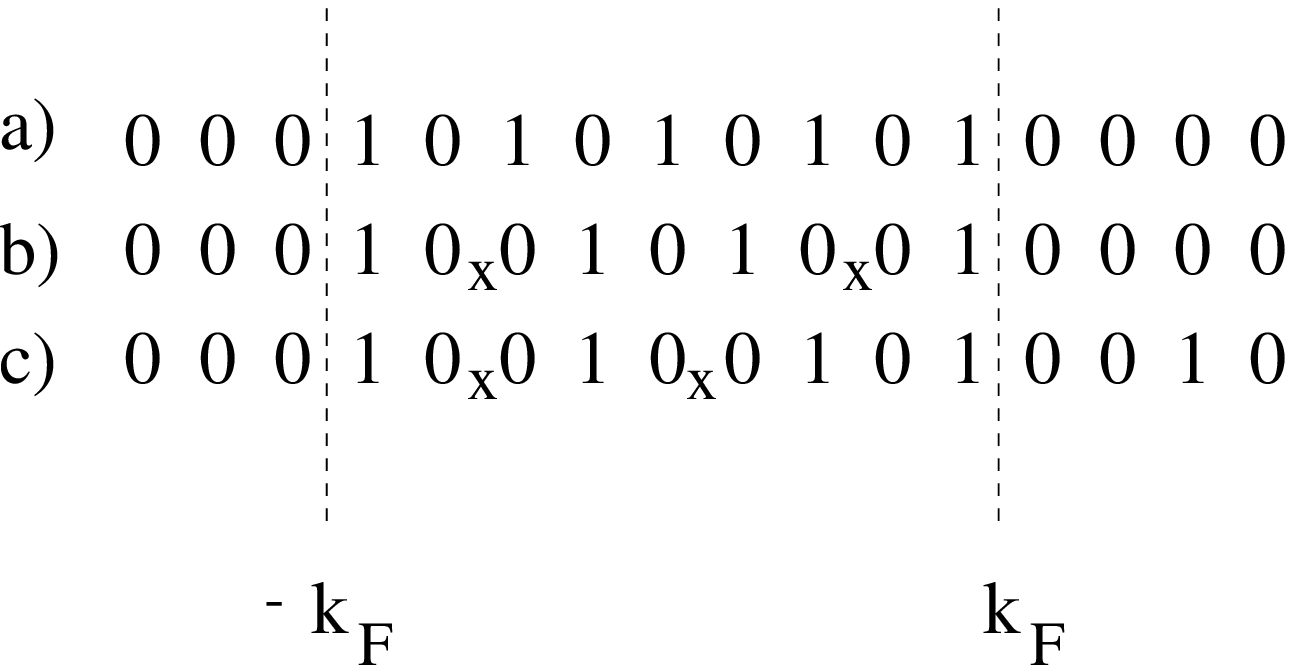}{8cm}
\figlabel\tabb
A typical configuration is shown in figure 1 b) for 
$\beta=2$. 
It is completely characterized by the positions of the $\be$ sequences
of $\be$ consecutive zeroes (denoted by 
crosses on the figure).  We shall see that these excitations
can be interpreted as $\beta$ holes whose dynamics is governed by a
Calogero-Sutherland hamiltonian at coupling $1/\be$.  In the case
of the Haldane-Shastry chain ($\beta=2$), this picture coincides
precisely with the low excitations found for the
XXX chain \faddeev . This is not surprising since the 
spectrum of the Haldane-Shastry chain can be obtained as a 
limit of the Bethe-Ansatz spectrum \Ha .

The density operator $\rho(x)=\Psi^\dagger(x) \Psi(x)$ can be
handled similarly.  $\Psi^\dagger \Psi(x)$ moves
a particle out of the Fermi-sea.  Thus, the intermediate 
states can be described with
$\beta$ holes propagating inside the Fermi sea and 1 particle
propagating outside (Fig. 1 c) .

The result for the Green function at $\beta=2$ is in accordance
with the expression previously obtained by Haldane and
Zirnbauer \Haldcorr .  
If $\beta> 2$, we get the expressions
conjectured by Haldane for the Green function
\Haldrev\ and which have been derived at equal time by
Forrester \Forres .  The density-density correlation was computed for the
values $\beta=1/2, 1, 2$ by Simons, Lee and Altshuler \alti\ and
conjectured by Haldane for all rational values of $\beta$ \Haldconj .
Here we prove the conjecture for $\beta$ positive integer. Forrester
has obtained the equal time limit of this result, his interpretation
however is different; he computes $<0|\Psi^\dagger \Psi^\dagger
\Psi \Psi|0>$ instead of $<0|\Psi^\dagger \Psi \Psi^\dagger \Psi|0>$.
Thus, he obtains $2\beta$ holes propagating in the intermediate 
states where we obtain $\beta$ holes and a particle.

Although there are conjectured expressions for the density-density
correlation for rational $\beta$ \Haldconj , we did not manage to
prove them using these techniques.  Also a generalisation to 
the $Su(N)$ spin chains and Calogero-Sutherland models \YB\
could be done along the same line, but then the properties of
the Jack polynomials studied in \stanley , \Macdo\ must be
generalized to the wave functions of these systems.

\newsec{Jack Symmetric Functions}

\subsec{Generalities.}

Before introducing Jack symmetric functions we shall define
a few quantities needed in the following \macdolivre . 
We denote a partition
$\{\lambda\}$ of the integer $\vert \lambda \vert$ by its parts
$\lambda_1\geq \lambda_2\geq... \geq0$ and such a partition can
be encoded in a Young tableau (Fig. 2). We define a partial
ordering by
$\{ \lambda \} \geq \{ \mu \}$ if $\vert \lambda \vert =
\vert \mu \vert$ and $\lambda_1+...+\lambda_i \geq
\mu_1+...+\mu_i$ for all $i\geq 1$.  
\fig{Young tableau}{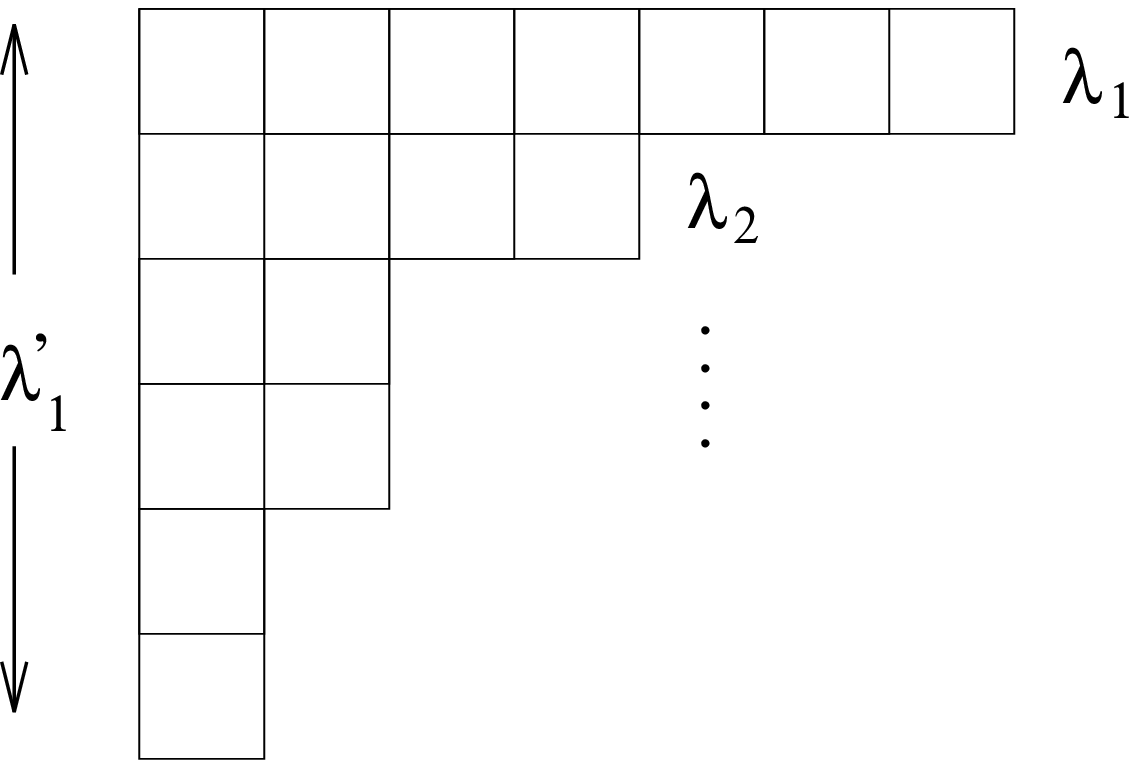}{6cm}
\figlabel\tabb
The conjugate of a partition is found by interchanging the
rows and columns in the corresponding Young tableau, it is
denoted by primed letters (eg. $\{ \lambda ' \} $).  The 
number of parts of a partition is called the length,
$l(\lambda )$, and is equal to $\lambda_1'$. The 
squares in the tableau are given the coordinates 
$(i,j), 1\leq j \leq \lambda_i $.  The coordinate $j$ 
increases towards the right starting from $1$,
the other coordinate, $i$, increases going downwards.
For each of these squares,
we define the arm-length, $a(s)$, leg-length, $l(s)$,
arm-colength, $a'(s)$, leg-colength, $l'(s)$, in the following
way~:
$$
\eqalign{& a(s)=\lambda_i-j , \ \  \ a'(s)=j-1\cr
& l(s)=\lambda'_j-i , \ \  \ l'(s)=i-1 .
}
$$
These are the number of squares right, down, left and up
a square $s\equiv (i,j)$ in a Young tableau.  The {\it hook-length}
of a square is $a(s)+l(s)+1$ and is the number of squares
below and to the right of it plus itself.  

We denote by $\Lambda _n$
the ring of  symmetric polynomials 
in $n$ indeterminates $z_1,...,z_n$.
There is a natural inclusion of $\Lambda_n$ in $\Lambda_{n+1}$
which consists in equating to zero the last variable
in $\Lambda_{n+1}$.  It is convenient to consider the formal
limit $\Lambda$  of $\Lambda_n$ where we let $n$ go to infinity. 
Natural basis of $\Lambda$ are given
by the elementary symmetric function in infinitely many
variables.
Let us here define the symmetric functions we use in the
following.

The power-sums are defined as~:
$$
p_i=\sum_k z_k^i
$$
and we can use them to obtain a basis of $\Lambda$ formed
by the power
sum symmetric functions~:
\eqn\powsum{
p_{\{\lambda\}}=p_{\lambda_1} p_{\lambda_2} ...
}
To
a partition $\{ \lambda \}$ we can also associate the monomial symmetric
function $m_{\{ \lambda \} }$ in these indeterminates.
It is the sum of all  distinct
permutations of the monomial $z_1^{\lambda_1} z_2^{\lambda_2}...$.
For example $m_{\{3,1\} }=\Sigma z_i^3 z_j$ for $i\neq j$.
If the number of indeterminates is less than the length of the
partition, the corresponding function $m_{\{\lambda \}}(z_i)$
is equal to zero.
A symmetric function $f\in \Lambda$ can be expressed in terms of
linear combinations of $m_{\{ \lambda \}}$'s.  
Another basis of $\Lambda_n$ is given by the Schur functions, 
$s_{\{\lambda\}}$, homogeneous polynomials of degree $\vert \lambda \vert$.
It is defined as~: 
\eqn\schur{
s_{\{\lambda \}}={\det (z_i^{\lambda_j+n-j})\over \det (z_i^{n-j})}
}
where the denominator is recognized to be the Vandermonde determinant.
It can be seen that the definition of $s_{\{\lambda \}}$ is 
compatible with the inclusion of $\Lambda_n$ into $\Lambda_{n+1}$
and therefore the $s_{\{\lambda \}}$ can be defined as a basis
of $\Lambda$.

\subsec{Definition and properties of Jack Symmetric functions.}

To a partition $\{\lambda \}$ having the part 1 with 
multiplicity $m_1$,
the part 2 with $m_2$, etc... we associate the number~:
$z_{\{\lambda\}}$
\eqn\zlam{
z_{\{\lambda\}}=1^{m_1} m_1! \ 2^{m_2} m_2! ...
}
We define a scalar product on $\Lambda$ given by~:
\eqn\normi{
<p_{\{\lambda\}},p_{\{\mu\}}>=\delta_{\{\lambda\},\{\mu\}}
z_{\{\lambda\}} \beta^{-l(\lambda )}
}

Then the Jack symmetric functions are 
symmetric functions in $\Lambda$,
uniquely defined by the following conditions~:
\eqn\jackdef{\eqalign{
&i) \ J_{\{ \lambda \} }(z_i; \beta) = m_{\{ \lambda \} }+ 
\sum_{\{ \mu \} < \{ \lambda \} }  v_{\{\mu\},\{ \lambda \} }(\beta) \
m_{\{ \mu \} } \cr
&ii) \ <J_{\{\lambda \} },J_{\{ \mu \} } > =0 \ \ , if \ \
\{ \lambda \} \neq \{ \mu \}
}}
where the value of $\beta$ is the same in the scalar product and
in the Jack polynomial.  For $\beta=1$ the Jack symmetric
functions coincide with the Schur functions.  It follows from
$i)$ that $J_{\{\lambda \}}(z_1,...,z_n;\beta)=0$ if
$n<l(\lambda )$.
Up to now there is no explicit expression for these
functions. The first ones are equal to~:
$$\eqalign{
J_{\{ 0 \} }&=1 \cr
J_{\{ 1^p \} }&= m_{\{ 1^p \} } \cr
J_{\{ 2 \} }&= m_{\{ 2 \} } + {2 \beta\over \beta+1} m_{\{ 1^2 \} } \cr
J_{\{ 3 \} } & = m_{\{ 3 \} } + {3\over \beta+2}
m_{\{ 2,1\} } + {6 \beta\over (\beta+1)(\beta+2)} m_{\{ 1^3 \} } \cr
J_{\{ 2,1\} } &= m_{\{ 2,1 \} }+ {6 \beta\over 2 \beta+1}
m_{\{ 1^3 \} }
} 
$$
When the number of indeterminates is  equal to $N$,
they are the eigenfunctions of a differential operator, $\tilde{H}(\be)$,
which is
a gauge transformed version of the Calogero-Sutherland hamiltonian~:
\eqn\gtrf{
\tilde{H}=\sum_{i=1}^N (z_i {\partial\over \partial z_i})^2 
+ \beta \sum_{i\neq j} {z_i+z_j \over z_i-z_j} z_i {\partial \over 
\partial z_i}
}
with the eigenvalue given by~:
\eqn\evalue{
e_{\{\lambda\}}(\beta)=\sum_i (\lambda_i+\beta (N-i))^2
}
This follows from the fact that $\tilde{H}$ is hermitian and triangular
in the $m_{\{\lambda \}}$ basis \Macdo .  It is then not difficult
to show that if one sets $z_k=e^{i 2 \pi x_k/L}$, the eigenfunctions
of $\hat{H}$, $\hat{P}$ defined in \imp\ and \halm\ are given by:
\eqn\efet{
\psi_{\{k\}}(x)=J_{\{\lambda\}}(z_i)\ \left( \prod_{i=1}^N z_i 
\right)^{q-(N-1)\beta /2} \times \prod_{i<j} (z_i-z_j)^\beta
}
The momenta $k_i$ defined in \estate\ 
are related to the partition $\{\lambda\}$ and to
$q$ (positive or negative integer) by the formula~:
\eqn\rela{
k_i=\lambda_{N-i+1}+q+\beta \(i-{N+1\over 2}\)
}
Using the explicit form of the
hamiltonian and the corresponding eigenvalue, it
is easy to see that~:
\eqn\homog{
\left( \prod_{i=1}^N z_i \right)
J_{\{\lambda \}}(z_1...z_N;\beta) 
= J_{\{ \lambda+1 \}}(z_1...z_N;\beta)
}
where by $\{\lambda+1\}$ we understand the partition in which
we have added one column of $N$ boxes.  So, to avoid double counting of states
in \efet , $\{\lambda \}$ must be
such that $l(\lambda )< N$ ($\lambda_N=0$). 

These functions have many interesting properties, one of them is the
duality which relates polynomials with coupling $\beta$ and $1/\beta$
\stanley\ \Gaudin :
\eqn\resui{
\prod_{i,j} (1+z_i w_j) = \sum_{\lambda } J_{\{ \lambda\} } (z_i,\beta)
J_{\{ \lambda'\} } (w_i,1/\beta) .
}

 To derive the results we  
rely on a few theorems which we now state. 
The first theorem is an extension of the constant term conjectures
of Dyson and Macdonald. We define a second norm, $(.,.)$ by~:
\eqn\normii{
(X,Y)_\beta =\int {d\theta_1\over 2 \pi} ... \int 
{d\theta_N\over 2 \pi} \ 
(\Delta(z_i)\overline{\Delta(z_i)})^\beta 
 \times \bar{X} Y.
}
where $\Delta(z_i)=\prod_{i<j} (z_i-z_j)$ and $z_k=e^{i \theta_k}$ is
a variable taking value on the unit circle.  We then have the
following theorem.

{\noindent\bf Theorem 1} (Macdonald \Macdo\ )
\eqn\theoi{
(J_{\{ \lambda \}},J_{\{ \mu  \} })_\beta= \delta_{\{\lambda\},\{\mu\}}
c_N \prod_{s\in \lambda}
{a'(s)+\beta (N-l'(s)) \over a'(s)+1+\beta(N-l'(s)-1) } \times
{h^*_\la(s) \over h_*^\la(s) }
}
where~: 
$$
c_N={(N\beta)! \over (\beta!)^N}
$$
and the quantities $h^*_\la$ and $h_*^\la$ are the 
upper hook-length and lower hook-length, defined as~:
\eqn\hook{\eqalign{
h^*_\la(s)&=\beta l(s)+a(s)+1 ,\cr
h_*^\la(s)&=\beta (l(s)+1)+a(s) 
}.}
The second theorem gives the value of the Jack polynomial when 
$p$ variables are equal to 1 and all other vanish.  

{\noindent \bf Theorem 2} (Stanley \stanley , Macdonald \Macdo\ ) 
\eqn\theoii{
J_{\{\lambda \} }(z_1=...=z_p=1,z_{p+1}=...=0;\beta)
=\prod_{s\in \lambda} {a'(s)+\beta (p -l'(s)) \over 
a(s)+\beta(l(s)+1 ) }
}

We can also, in parallel to Schur functions, define the 
notion of {\it skew Jack polynomial}, $J_{\{\lambda/\mu\}}(
z_i;\beta)$, through~:
\eqn\ske{
<J_{\{\lambda/\mu\}},J_{\{\nu\}} >\ =\ <J_{\{\lambda\}},J_{\{\mu\}}
J_{\{\nu\}} >.
}
Using these functions, given 2 sets of variables $z_i$ and
$w_i$, one can  show that
\stanley  \Macdo ~:
\eqn\separ{
J_{\{\lambda \}}(z_i,w_i;\beta)=\sum_{\{\nu\}}
J_{\{\nu\}}(z_i;\beta) J_{\{\lambda/\nu\}}(w_i;\beta)
}
In the specialized form where the $w_i$'s are 
represented by only one variable, $u$, we have :
\eqn\skef{
J_{\{\lambda \}}(u,z_i;\beta)=\sum_{\{\nu\}}
J_{\{\lambda/\nu\}}(u;\beta)
J_{\{\nu\}}(z_i;\beta)
}
with the skew function different from zero only when $\{\lambda-\nu\}$ is
a horizontal strip \stanley , i.e., when all $\lambda_i'-\nu_i'=0,1$. In
other words, the diagram $\{\lambda \}$ contains that of
$\{\nu \}$ ($\lambda_i\geq \nu_i$ for all $i$) and the
difference denoted $\{\lambda -\nu\}$ has at most one box in
each columns.
If $\{\lambda -\nu\}$ is a horizontal strip, we define $C_{\{\lambda/\nu\}}$
to be the set of the columns for which 
$\lambda_i'-\nu_i'=0$. 
An explicit expression for 
$ J_{\{\lambda/\nu\}}(u;\beta)$ is given by :

{\noindent \bf Theorem 3}(Stanley \stanley\ )
\eqn\skefi{
J_{\{\lambda/\nu\}}(u;\beta)=u^{\vert \lambda-\nu \vert}
\prod_{s\in C_{\{\lambda/\nu\}}}
\left({h^*(s)\over h_*(s)}\right)_\lambda \left( {h_*(s)\over h^*(s)}
\right)_\nu 
}
where the notation $( ...)_\lambda$ means that we evaluate the
upper and lower hook-length with respect to the partition $\{
\lambda \}$.

\newsec{Retarded Green function.}

The retarded single
particle Green function, $<0|\Psi^\dagger(x,t)
\Psi(0,0)|0>$, can be simply computed at integer coupling.
From now on we will use variables $z_k=e^{2i \pi x_k/L}$.
In these variables the ground state is given by :
\eqn\gro{
|0>_{N+1}=\Delta_{N+1}^\beta (z_i) \prod_i z_i^{-\beta N/2}.
}
When acting upon it with the operator $\Psi(0,0)$ we\foot{
We put $N+1$ particles in the ground state to ease the
forthcoming notation.} fix the variable $z_{N+1}$ to be
equal to 1 and obtain :
\eqn\wavefunc{
\Psi(0,0) \vert 0 >_{N+1}=\prod_{i=1}^{N} (z_i-1)^\beta \prod_{i=1}^N 
z_i^{-\beta/2} |0>_N .
}
Using :
\eqn\enemom{
\Psi^\dagger(x,t)=e^{i(\hat{H} t-\hat{P}x)} \Psi^\dagger(0,0)
e^{-i(\hat{H}t-\hat{P}x)}
}
the correlation we are seeking can be written as :
\eqn\denscorr{
\eqalign{
<0\vert& \Psi^\dagger (x,t) \Psi(0,0) \vert 0>\ =\cr 
&c_{N+1}(\beta)^{-1} <0\vert \prod_{i=1}^{N} z_i^{\beta/2}
(1-\bar{z_i})^\beta 
e^{-i((\hat{H}-E_0)t+(\hat{P}-P_0)x)} \prod_{i=1}^{N} z_i^{-\beta/2}
(1-z_i)^\beta \vert 0>_N\cr
}}
The notation $<0|...|0>_N$ means
that we integrate over the
angles $\theta_i=2 \pi x_i/L$
with $i=1,..N$, which amounts to compute 
the second norm defined in the last section ($c_{N+1}$ is simply 
the normalization $<0|0>_{N+1}$).

It can be done by expanding the product on the 
Jack symmetric functions using \resui\ where we set
$w_j=-1$ for $1\leq j \leq \beta$ :
\eqn\expan{\eqalign{
\prod_{i=1}^{N}(1-z_i)^\beta &=\sum_{\{\lambda\} \atop
l(\lambda')\leq \beta, l(\lambda)\leq N}
J_{\{\lambda\}} (z_i;\beta) \
J_{\{\lambda'\}} (w_1=...=w_\beta=-1,0...;{1\over \beta}) \cr
&=\sum_{\{\lambda\} \atop l(\lambda ')\leq \beta, l(\lambda)\leq N}
(-1)^{\vert \lambda \vert}
J_{\{\lambda\}}(z_i;\beta) \ 
J_{\{\lambda'\}}(w_1=...=w_\beta=1;{1\over \beta})
}
}
The length of $\{\lambda \}$ and $\{\lambda '\}$, $l(\lambda)$ and
$l(\lambda')$ are respectively less or equal than $N$ and $\beta$
because the number of variables in $J_{\{\lambda \}}$ and
$J_{\{\lambda '\}}$ is respectively equal to $N$ and $\beta$.
As shown before, the wave functions defined by \expan\ 
are eigenstates of the hamiltonian and the momentum operators. Let
us denote by $E_{\{\lambda \}}$ the eigenvalue of $\hat{H}-E_0$
where $E_0$ is the ground state energy (with $N+1$ particles)
and by $Q_{\{\lambda \}}$ the eigenvalue
of $\hat{P}$.  
Expanding both products in \denscorr\ 
in that way and using the fact that
the Jack symmetric functions are orthogonal to each other \normii , 
we obtain : 
\eqn\densfinal{\eqalign{
<0\vert \Psi^\dagger (x,t) &\Psi(0,0) \vert 0>\ =\cr &
\sum_{\{\lambda\}\atop l(\lambda')\leq \beta, \ l(\lambda)\leq N} 
e^{-i E_{\{\lambda \}} t+i Q_{\{\lambda \}} x} 
N_{\{\lambda \}} \left( J_{\{\lambda'\}}(w_1=...=w_\beta=1,0...
;{1\over \beta}) \right)^2.
}}
It remains to  evaluate  the norm, $N_{\{\lambda \}}$ and
the Jack polynomial $J_{\{\lambda'\}}$ at $w_1=w_2=...=w_{\beta}=1$ 
.  These two quantities derive
from the theorems 1 and 2.  As an example we show on Fig.3 
how to evaluate $J_{\{\lambda '\}}(1,1,0,0,... ; 1/2)$ for
$\beta=2$, that is~:
\eqn\betii{
\prod_{s\in \{ \lambda'\}} {j-i/2+1/2 \over \lambda_i'-j-i/2
+\lambda_j/2+1/2}
}
where the partition $\{\lambda '\}$ has only two rows.
\fig{Evaluation of $J_{\{\lambda'\}}(1,1,0...;1/2)$ }{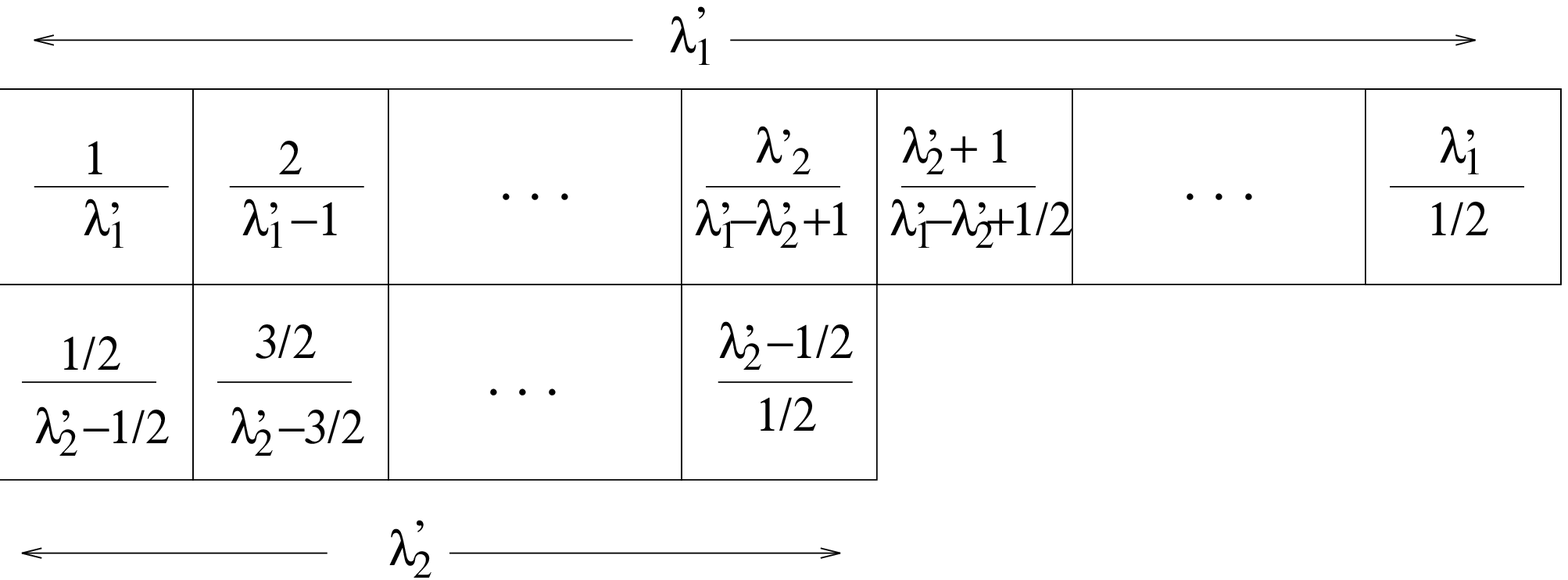}{9cm}
\figlabel\tabb
The result is~:
\eqn\resulbe{
J_{\{\lambda_1',\lambda_2'\}}(1,1,0,...;1/2)=\Gamma (1/2) 
{\Gamma (\lambda_1'-\lambda_2'+1) \over \Gamma(\lambda_1'-\lambda_2'
+1/2) } .
}
The generalisation of this expressions 
to an arbitrary value of $\beta$
and the norms are given 
in the appendix A.  We give
here the final result for the correlation function \densfinal :
\eqn\finitn{\eqalign{
\eta(x,&t,0,0)={\beta! \over N+1}\sum_{0\leq \lambda_\beta'\leq ...
\leq \lambda_1'\leq N}\  
\prod_{i<j} {\Gamma(\lambda_i'-\lambda_j'+{j-i+1\over \beta})
\Gamma(\lambda_i'-\lambda_j'+1+{j-i\over \beta}) \over
\Gamma(\lambda_i'-\lambda_j'+{j-i\over \beta}) \Gamma(\lambda_i'
-\lambda_j'+1+{j-i-1\over \beta})} \cr \noalign{\vskip3pt}&
\times \left(
\prod_{i=1}^\beta {\Gamma(1+{1\over \beta}) 
\Gamma(\lambda_i'+1+{1-i\over \beta}) \Gamma(N-\lambda_i'+i/\beta)
\over \Gamma(i/\beta)^2 \Gamma(\lambda_i'+2-i/\beta)
\Gamma(N-\lambda_i'+1+{i-1\over \beta}) } \right)
 e^{-i E_{\{\lambda \}}t
+i Q_{\{\lambda \}} x}
}
}
The values of the energy and momentum are :
\eqn\enmom{\eqalign{
E_{\{\lambda\}}&=\sum_{i=1}^N \left( \lambda_i+\beta({N+2\over 2}
-i)\right)^2-\sum_{i=1}^{N+1} \beta^2 \left( {N+2\over 2}-1\right)^2
\cr &
={2\pi^2\be\over L^2}\left(\sum_{i=1}^\be\la_i'\(N+1-\la_i'+\frac{2
i-1}{\be}\)-\beta {N^2\over 4}\right)
}}
and~:
\eqn\immmm{
Q_{\{\lambda\}}={2 \pi^2\over L} \sum_i \left( \lambda_i' -{N\over 2}
\right)
}

In the thermodynamic limit, $N\rightarrow \infty$ , $L\rightarrow \infty$,
we define the variables :
\eqn\varch{
v_i={2\over N} (\lambda_i'-{N\over 2})
}
and use the Stirling formula :
\eqn\stir{
\Gamma(N+1)\vert_{N\rightarrow \infty} \simeq \sqrt{2\pi}
N^{N+1/2} \ e^{-N}
}
We also write the sums as integrals and remove the order on partitions :
\eqn\sint{
{2\over N} \sum_{0\leq\lambda_i'\leq \lambda_{i-1}'}
\rightarrow \int_{-1}^{v_{i-1}} dv_i
}
We finally obtain :
\eqn\final{\eqalign{
<0|\Psi^\dagger&(x,t) \Psi(0,0)|0>= \cr &
{\rho \over 2} \left(
\prod_{i=1}^\beta \int_{-1}^1 dv_i \right) e^{i(Qx-Et)}
\prod_{i=1}^\beta {\Gamma(1+{1\over \beta})\over
\Gamma(i/\beta)^2} \ (1-v_i^2)^{-1+1/\beta} \times 
 \prod_{i<j} \vert v_i-v_j\vert^{2\over \beta} 
}}
In this limit the energy and momentum eigenvalues 
are given by :
\eqn\apphalm{
E={-\beta \rho^2 \pi^2 \over 2}
\sum_{i=1}^\beta v_i^2.
}
\eqn\mommm{
Q=\rho \pi \sum_{i=1}^{\beta}v_i.
}

\newsec{Density-Density Correlation}

The
density-density correlation function 
can be expressed as :
\eqn\densdens{\eqalign{
{<0\vert \rho(x,t) \rho(0,0) \vert 0>\over <0\vert 0>}
&=\sum_{\alpha } {1\over N_{\{0\}}(\beta)
N_{\alpha }(\beta) }
<0|\rho(x,t)|\alpha ><\alpha |\rho(0,0)|0>
\cr
&=\sum_{\alpha } {e^{i(Q_{\alpha } x-E_{\alpha } t)}\over
N_{\{0\}}(\beta) N_{\alpha }(\beta)} 
\vert <0\vert \rho(0,0)\vert \alpha >\vert^2.
}}
where $|\alpha >$ stands for a complete set of states characterized
by their momenta $k_i$ as explained in the introduction.
Let us first consider the intermediate states 
which have all the momenta $k_i\geq -k_F$
where $k_F$ is the Fermi momentum.
They can be labeled by the partition of the Jack polynomial,
$\{\lambda \}$.  

Consider the ground state with $N+1$ particles, then
the density operator $\rho(0,0)=\sum_i \delta(x_i)$
has a form factor equal to :
\eqn\ffdens{\eqalign{
<0|\rho(0,0)|\{\lambda \}>=&
\int {d\theta_1\over 2 \pi} ... {d\theta_N\over 2 \pi}
(\Delta(z_i) \overline{\Delta(z_i)})^\beta
\times\cr \times & 
\prod_{i=1}^N (1-z_i)^\beta (1-\bar{z_i})^\beta 
\ J_{\{\lambda \}}(1,z_1,...,z_N; \beta)
}}
Since we are working on the unit circle, $z_i=1/\bar{z_i}$ and we
can rewrite:
\eqn\rewr{
\prod_i (1-z_i)^\beta(1-\bar{z_i})^\beta =(-1)^{\beta N}
\prod_i {(1-z_i)^{2 \beta} \over z_i^{\beta}}
}
Expanding the product on the numerator with \resui\ and
using \skef\ to separate the first variable in the
Jack polynomial we get the expression :
\eqn\interm{\eqalign{
<0|&\rho(0,0)|\{\lambda \}>=\int {d\theta_1\over 2\pi} ...
{d\theta_N\over 2 \pi}
(\Delta (z_i) \overline{\Delta(z_i)})^\beta  
\cr \noalign{\vskip3pt} &\sum_{\{\mu\} ,\{\nu \}} (-1)^{\vert \mu 
\vert+\beta N} 
J_{\{\mu\}}(z_i;\beta) J_{\{\nu \}}(z_i;\beta) \times \cr 
\noalign{\vskip2pt}& \times
J_{\{\mu'\}}(w_1=...w_{2\beta}=1,0...;1/\beta)
J_{\{\lambda/\nu\}} (1;\beta) \prod_i z_i^{\beta}.
}}
The product over all variables $z_i$ can be incorporated into
one of the Jack polynomials using \homog ; after integration
we obtain the final expression :
\eqn\ddfin{\eqalign{
<0|&\rho(0,0)|\{\lambda\}>=\cr &\sum_{\{\mu\}}
N_{\{\mu\}}(\beta) (-1)^{\vert \mu \vert} J_{\{(\mu+\beta)'
\}}(w_1=...=w_{2\beta}=1,0...;{1\over \beta})
J_{\{\lambda/ \mu\}}(1;\beta)
}}	
where the partitions $\{\mu+2\}$ and $\{\lambda\}$ are illustrated
in the case $\beta=2$ in figure 4.
\fig{Example of partitions $\{\mu+2\}$ and $\{\lambda \}$ for 
$\beta=2$}{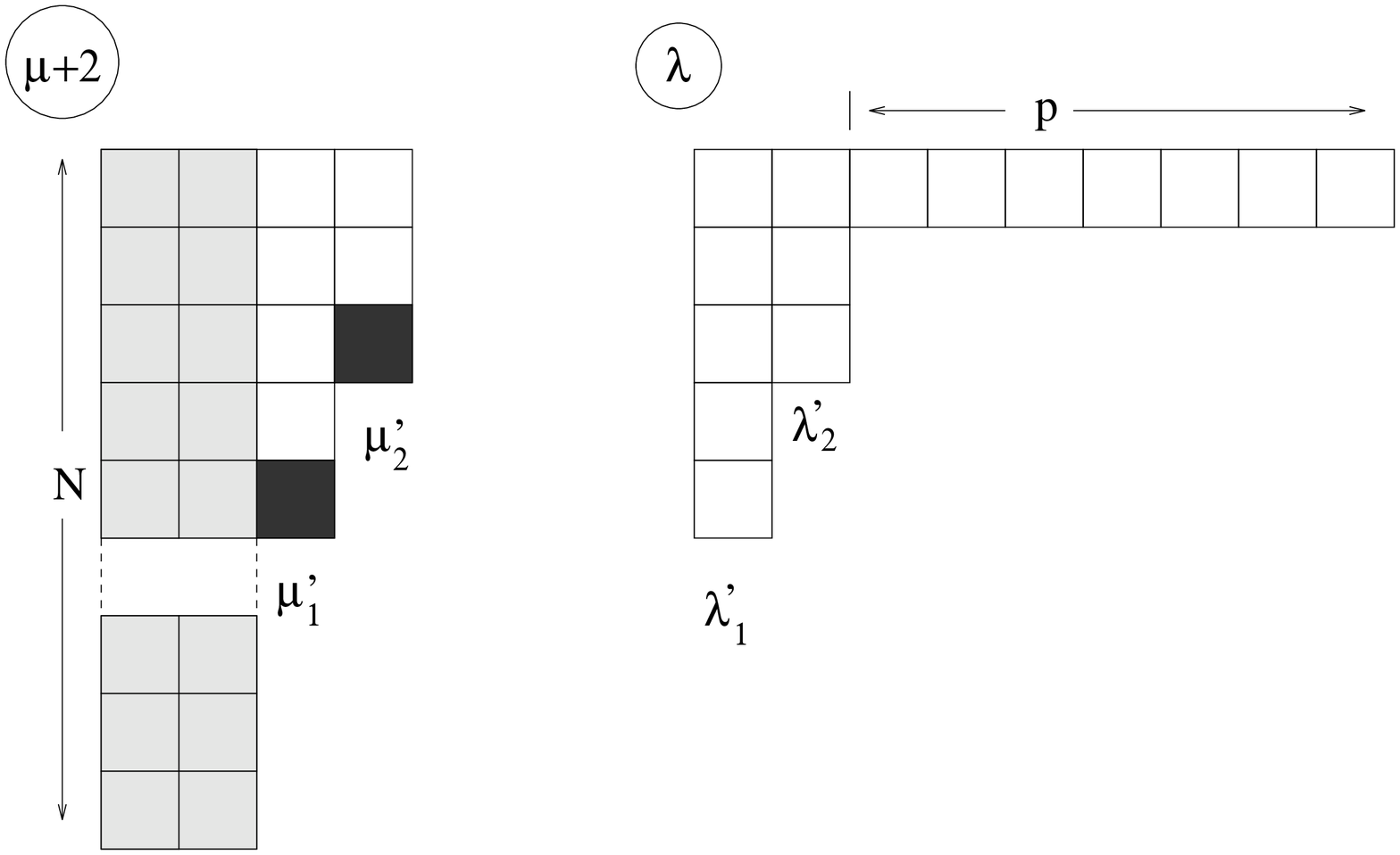}{10cm}
\figlabel\tabb
In \ddfin\ the
evaluation of the norm, and that of the Jack polynomial
at 1 can be done in exactly the same manner as precedently.  
We need to identify which partitions $\{\mu\}$ contribute
to the sum and for these partitions evaluate 
$J_{\{\lambda/ \mu\}}(1;\beta)$. 
It follows from theorem 3, that all partitions 
$\{\mu\}$ are included in $\{\lambda\}$
and differ from it by at most
one box at the bottom of its columns, that is
all possibilities $\lambda_i'-\mu_i'=0,1$.

From the expansion of \rewr\ it follows that only 
partitions $\{\mu \}$ with $\beta$ columns contribute
to the sum.  Then the non vanishing contributions from
partitions $\{\lambda \}$ are those with $\beta$ legs
and 1 arm of length $\lambda_1=p+\beta$.

For example, when $\beta=2$ (Fig. 4) we have four
possibilities, a) when both dark shaded boxes are present,
b), c) when one of the dark shaded box is present, d) when
no dark shaded boxes is present.  Then theorem 3 states 
that the corresponding value of $J_{\{\lambda /\mu\} }
(1;\beta)$ is the product over squares of 
columns having shaded boxes of the following quantity~:
\eqn\eg{
\left( {h^*(s)\over h_*(s)} \right)_\lambda 
\left( {h_*(s)\over h^*(s)} \right)_\mu
}
where the upper and lower hook-length are taken with respect 
to  the partition $\{\lambda \}$ or $\{\mu \}$ respectively.

Let us denote by $I$ the ensemble of columns of
$\{\mu\}$ equal to the corresponding ones in 
$\{\lambda\}$ .  Then the sum over partitions $\{\mu\}$
can be replaced by a sum over the ensembles $I$  
.  Since
the partition $\{\mu\}$ must be included in
the partition $\{\lambda\}$ let us factor out of the
sum the terms for which all parts $\mu_i'=\lambda_i'-1$;
we obtain the following expression for the form factor :
\eqn\monstre{
\eqalign{
<0|\rho(0,0)|\{\lambda\}>&=
c_N(\beta) 
\left( \prod_{i=1}^\beta {\Gamma(N+1+{i-1\over \beta})\over
\Gamma(1+i/\beta) \Gamma(N+i/\beta)} \right)
(-1)^{\vert \lambda \vert -\beta-p}
\cr \noalign{\vskip3pt}
& \times \left(\prod_{i=1}^\beta {\Gamma(N-\lambda_i'+2+i/\beta)
\Gamma(\lambda_i'-{i-1\over \beta})\over
\Gamma(N-\lambda_i'+2+{i-1\over \beta}) \Gamma(\lambda_i'+1-i/\beta)}
\right) \cr \noalign{\vskip3pt}
& \times 
\prod_{i<j} {\Gamma(\lambda_i'-\lambda_j'+1+{j-i\over \beta})
\over \Gamma(\lambda_i'-\lambda_j'+1+{j-i-1\over \beta}) }
\times \sum_{\{\mu\}} Q(\{\mu\})
}}
with the sum over the  partitions $\Sigma_{\{\mu\}} Q(\{\mu\})$ 
being equal to the following sum over  the ensembles $I$ :
\eqn\txx{\sum_I (-1)^{card(I)}
\prod_{i\in I} {p-i+1+\beta \lambda_i' \over
p-i+\beta (\lambda_i'+1)}\times 
{N-\lambda_i'+1+{i-1\over \beta} \over
N-\lambda_i'+1+i/\beta} \times \prod_{j\neq I}
{\lambda_i'-\lambda_j'+{j-i+1\over \beta} \over
\lambda_i'-\lambda_j'+{j-i\over \beta} }
.}
We evaluate this sum in appendix B, the result is :
\eqn\sasa{
{ (\Sigma_i \lambda_i' +p) \ \beta ! \ {\beta N+p+2 \beta-1 \choose \beta-1}
 \ (\beta-1)!
\over
\beta^\beta
\prod_{i=1}^\beta (p-i+\beta (\lambda_i'+1)) (N-\lambda_i'+1+i/\beta)}
}
where in the general case $p$ is $\lambda_1-\beta$ (see Fig. 4). 
The norm $N_{\{\lambda \}}$ needed to evaluate \densdens 
is  given in the appendix A.
We also need to take into account the
states for which the particle has a momentum less than $-k_F$,
which amounts to replace $e^{iQ_{\{\lambda \}}x}$ by
$2 \cos (Q_{\{\lambda \}}x)$) in the final expression.
The result is :
\eqn\ff{\eqalign{&\ \ \ g_{N,\beta}(x,t)=
{2 \beta!\over (N+1)^2}\  \prod_{i=1}^\beta {\Gamma(1+1/\beta)
\over \Gamma^2(i/\beta)}
\sum_{0\leq \lambda_\beta'\leq ... \leq \lambda_1'\leq N 
\atop 0\leq p \leq \infty} 
\cos(Q_{\{\lambda \}} x) e^{-iE_{\{\lambda \}}t} \cr &
\left( \prod_{i=1}^\beta {\Gamma(N-\lambda_i'+1+i/\beta) 
\Gamma(\lambda_i'+{1-i\over \beta}) \over
\Gamma(N-\lambda_i'+2+{i-1\over \beta}) \Gamma(\lambda_i'+1-i/\beta)}
\right) 
\prod_{i<j} {\Gamma(\lambda_i'-\lambda_j'+1+{j-i\over \beta}) 
\Gamma(\lambda_i'-\lambda_j'+{j-i\over \beta}) \over
\Gamma(\lambda_i'-\lambda_j'+1+{j-i-1\over \beta}) 
\Gamma(\lambda_i'-\lambda_j'+{j-i+1\over \beta}) }
\cr \noalign{\vskip5pt} &\ \ \ \  \times
{\Gamma(p+\beta) \Gamma(\beta N+2 \beta+p) 
(p+\Sigma \lambda_i')^2  \over \Gamma(p+1) \Gamma(\beta N+\beta+p+1)
\prod_i (p+\beta \lambda_i'+\beta-i) (p+\beta \lambda_i'-i+1) }
}}
with the explicit form of the energy and momentum given by :
\eqn\expli{\eqalign{
E_{\{\lambda\}}=&{2 \pi^2 \over L^2} \left(
\left[ \sum_{i=1}^\beta \lambda_\beta' \beta (N+1+{2 i-1\over \beta}
-\lambda_\beta')\right] +p^2+\beta(N+2)p \right)  
\cr 
Q_{\{\lambda \}}=&{2\pi \over L} \vert \lambda \vert.
}}
In the thermodynamic limit though everything
simplifies drastically.
As in the last section, we use
Stirling's formula and the change of variables :
\eqn\chvar{
{p\over \be N}=-{ w+1 \over 2} , \quad {\lambda_i'\over N}
={v_i+1\over 2}
}
and the following limits for the energy and momentum :
\eqn\edenden{
E={\beta \pi^2 \rho^2 \over 2} \left(
-\sum_i v_i^2 + \beta w^2 \right)
}
\eqn\pdenden{
Q=\pi \rho ( \sum_i v_i - \beta w)
}
We then find for the density-density correlation function :
\eqn\therlim{\eqalign{
A(\beta) \int_{-\infty}^{-1}& dw\int_{-1}^1 dv_1 ... \int_{-1}^1 dv_\beta \
(w^2-1)^{\beta-1} \prod_{i=1}^\beta (1-v_i^2)^{-1+1/\beta} \times \cr
\noalign{\vskip2pt} &\times \prod_{i<j} \vert v_i-v_j\vert^{2/\beta}  \times
{(\Sigma_i v_i-\beta w)^2 \over \prod_{i=1}^\beta (v_i-w)^2 }
 e^{-iEt} \cos (Q x)
 }
}
with the constant~:
\eqn\const{
A(\beta)={\rho^2 \over 2 \beta} \prod_{i=1}^\be
{\Gamma(1+1/\beta)\over \Gamma(i/\beta)^2 }.
}

{\noindent \it Note:} We have just received a paper by 
Polychronakos and Minahan \pm\ who also conjecture this formula
and also a paper by Z.N.C. Ha \ha\ who announces the same results.

\newsec{Acknowledgements}

We wish to thank D. Bernard and M. Gaudin for valuable discussions and
F.D.M. Haldane for sending us his conjectures prior
to publication.
V. Pasquier is grateful to the Erwin Schr\"odinger
International Institute for Mathematical Physics where part of this
work was done and to F. Essler and H. Grosse for discussions.
F. Lesage is supported by a Canadian NSERC 67 scholarship.

\appendix{A}{Norms and Evaluations of Jack polynomials at specific values.}

We list in this appendix the results found for the norms and
evaluation of the Jack polynomials at specialized values.

In section 3 we need to evaluate the Jack polynomial defined on a
partition $\{\lambda\}=\{\lambda_1',...,\lambda_\beta'\}$ (where
we defined the partition by its conjugate) at 
$z_1=...=z_\beta=1$.  The result is :
\eqn\pigen{
J_{\{\la'\}}(z_1=..=z_{\beta}=1,0,0,...;{1\over \beta})=
\prod_{i=0}^{\beta-1} {\Gamma (1/\beta) \over
\Gamma (1-i/\beta) } \times
\prod_{i<j} {\Gamma (\lambda_i'-\lambda_j'+{j-i+1\over \beta})
\over \Gamma (\lambda_i'-\lambda_j'+{j-i\over \beta})}
}
We also need the norm of the same partition,
$N_{\{\lambda\}}(\beta)$, it is given by :
\eqn\norme{\eqalign{
N_{\{\lambda\}}(\beta)=c_N(\beta) \prod_{i=1}^\beta&
{\Gamma(N+1+{i-1\over \beta}) \over \Gamma(1/\beta) 
\Gamma(N+i/\beta)}
{\Gamma (\lambda_i'
+1+{1-i\over \beta} )\over 
\Gamma(\lambda_i'+2-i/\beta) }{\Gamma(N-\lambda_i'+i/\beta) \over
\Gamma(N-\lambda_i'+1+{i-1\over \beta})}
\times 
\cr \noalign{\vskip3pt}& \times \prod_{i<j}
{\Gamma (\lambda_i'-\lambda_j'+{j-i\over \beta}) \Gamma(
\lambda_i'-\lambda_j'+1+{j-i\over \beta}) \over
\Gamma(\lambda_i'-\lambda_j'+{j-i+1\over \beta}) 
\Gamma(\lambda_i'-\lambda_j'+1+{j-i-1\over \beta})}
}}

In section 4 we need to evaluate the norm of two Jack polynomials
on a partition of type $\{\lambda \}=\{p;\lambda_1',...,\lambda_\beta'\}$
where the $\lambda_i'$'s denote the first $\beta$ legs of the
partition and $p$ denote $\lambda_1-\beta$.  
The norm is found to be :
\eqn\nlam{\eqalign{
c_{N+1}(\beta) \prod_{i=1}^\beta &
{\Gamma(N+2+{i-1\over \beta}) \Gamma(N-\lambda_i'+{i\over\beta}+1) 
\Gamma(\lambda_i'+{1-i\over \beta}) (p+\beta \lambda_i'-i+1) \over
\Gamma(N+{i\over\beta}+1) \Gamma(N-\lambda_i'+2+{i-1\over \beta})
\Gamma(\lambda_i'+{\beta-i\over\beta}) \Gamma({1\over \beta}) 
(p+\beta \lambda_i'-i+\beta) } 
\cr \noalign{\vskip5pt} &\times 
{\Gamma(p+1) \Gamma(\beta) \Gamma(\beta (N+1)+1) \Gamma(p+\beta (N+2))
\over \Gamma(p+\beta) \Gamma(\beta (N+2)) \Gamma(p+\beta (N+1)+1) }
\cr \noalign{\vskip5pt} & \times \prod_{i<j}
{\Gamma(\lambda_i'-\lambda_j'+{j-i\over \beta}) \Gamma(\lambda_i'
-\lambda_j'+1+{j-i\over \beta}) \over
\Gamma(\lambda_i'-\lambda_j'+{j-i+1\over \beta}) 
\Gamma(\lambda_i'-\lambda_j'+1+{j-i-1\over \beta}) }
}}
Another quantity we used to calculate the density-density correlation
function is the polynomial associated to the partition with $\be$ arms
of length $N$ and $\be$ arms of length $\la_1',\ldots,\la_\be'$
evaluated at $z_1=...=z_{2\be}=1$. The result is :
\eqn\Nlam{\eqalign{
&J_{\{(\be+\la)'\}}(z_1=...=z_{2\be}=1,0,...;{1\over\be}) \cr
&= \prod_{i=1}^\be {\be\over i}\(N-\la_i'+{i \over \be}\)\times
J_{\{\la'\}}(z_1=..=z_{\beta}=1,0,0,...;{1\over \beta})
}}
	
\appendix{B}{Summation of \txx .}

We wish to evaluate the following expression,
\eqn\expeval{\sum_I (-1)^{card(I)}
\prod_{i\in I} {p-i+1+\beta \lambda_i' \over
p-i+\beta (\lambda_i'+1)}\times 
{N-\lambda_i'+1+{i-1\over \beta} \over
N-\lambda_i'+1+i/\beta} \times \prod_{j\not\in I}
{\lambda_i'-\lambda_j'+{j-i+1\over \beta} \over
\lambda_i'-\lambda_j'+{j-i\over \beta} }
.}
where $I \subset \{1,\ldots,\be\}$ and $J$ is the complement of $I$.
Changing variables and writing $u_i=\lambda_i'-i/\beta+p/\beta$
and $y=N+p/\beta+1$ we obtain the simpler expression~:
\eqn\simple{
\sum_I (-1)^{card(I)} \prod_{i\in I}
{u_i+1/\beta\over u_i+1} \times {y-u_i-1/\beta\over y-u_i}
\times \prod_{i\in I \atop j \in J} {u_i-u_j+1/\beta \over
u_i-u_j}
}
By considering the 2 terms with $i\in I$, $j\in J$ and
$i \in J$ and $j\in I$, all other indices being fixed, one
sees that this expression has no pole at $u_i=u_j$.  It is
symmetric under permutations of the indices.  Thus, putting everything
on the same denominator, the numerator is a symmetric
polynomial of degree at most two in each variables, $u_i$.  
By considering the two terms with $i\in I$ and $i\in J$ all other
indices being kept fixed, one sees that the numerator is in fact
of degree one in each variables $u_i$. It  is
therefore determined by the $\beta+1$ coefficients of the symmetric
polynomials $m_{\{1^k\}}$, $0\leq k\leq \beta$.  If we specialize
the variables to $u_i=u-i/\beta$, these coefficients are 
determined by considering the polynomial in $u$ at the numerator.
In that case the only terms 
contributing to the sum are those for which the first 
$p$ terms are in the ensemble $I$ and the $\beta - p$ last terms
are in the complement $J$.  The sum is then written as :
\eqn\wir{
\sum_{p=0}^\beta (-1)^p {\beta \choose p} 
\prod_{j=1}^p {u+{1-j\over \beta} \over
u+1-j/\beta} \times {y-u+{j-1\over \beta}\over
y-u+j/\beta} 
}
making use of the identity\foot{
We denote $(a)_\beta={\Gamma(a+\beta)\over \Gamma(a)}$} \tables\ :
\eqn\identite{\eqalign{
&\sum_{p=0}^\beta {(-\beta)_p (c)_p (a+b-c+\beta-2)_p \over
p! (a)_p (b)_p} =\cr &
{(a-c-1)_\beta (b-c-1)_\beta \over (a)_\beta (b)_\beta (a-c-1)
(b-c-1)} [(a-c-1)(b-c-1)+\beta (a+b-c+\beta-2)]
}}
\wir\ becomes :
\eqn\hres{
{(\beta y+1)_\beta \beta! (\beta u) \over 
\prod_{i=1}^\beta (\beta (y-u)+i) (\beta u+\beta-i) }
}
We finally  find the sum \expeval\ to be~:
\eqn\firesul{
{ (\Sigma_i \lambda_i' +p) \ \beta ! \ {\beta N+p+2 \beta-1 \choose \beta-1}
 \ (\beta-1)!
\over
\beta^\beta
\prod_{i=1}^\beta (p-i+\beta (\lambda_i'+1)) (N-\lambda_i'+1+i/\beta)}
}

\listrefs
\end